\documentclass[twocolumn,pre,showpacs,floatfix]{revtex4}
\usepackage{graphicx}

\begin{document}

\title{Molecular Discreteness in Reaction-Diffusion Systems Yields\\
Steady States Not Seen in the Continuum Limit}
\author{Yuichi Togashi}
\email{togashi@complex.c.u-tokyo.ac.jp}
\author{Kunihiko Kaneko}
\affiliation{Department of Basic Science, School of Arts and Sciences, The
University of Tokyo,
Komaba, Meguro, Tokyo 153-8902, Japan}
\date{August 5, 2004}

\begin{abstract}
We investigate the effects of spatial discreteness of molecules
in reaction-diffusion systems.
It is found that discreteness within the so called Kuramoto length
can lead to a localization of molecules, resulting in novel steady states
that do not exist in the continuous case.  These novel states  are
analyzed theoretically as the fixed points of accelerated localized reactions,
an approach that was verified to be in good agreement with
stochastic particle simulations.
The relevance of this discreteness-induced state
to biological intracellular processes is discussed.
\end{abstract}


\maketitle

% --- figures

\begin{figure*}
\begin{center}
\includegraphics[width=58mm]{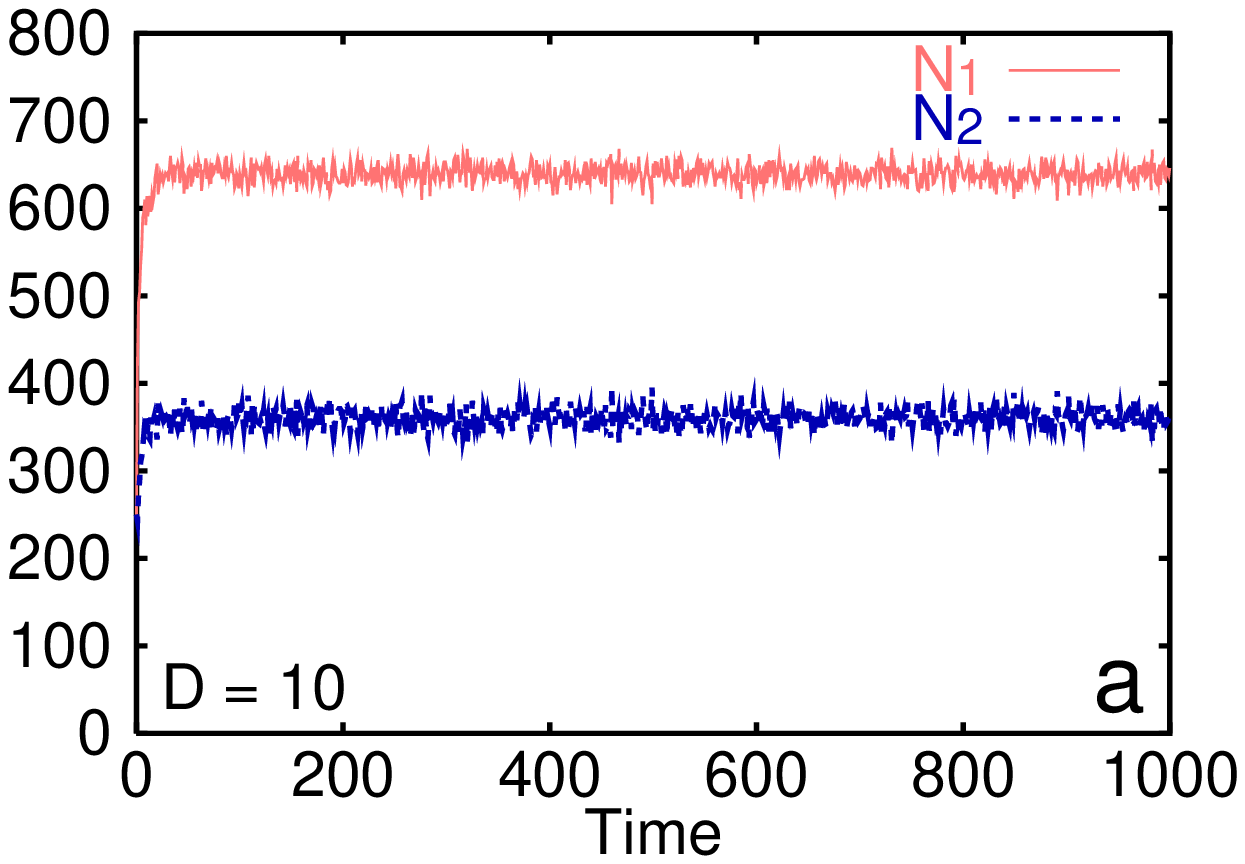}
\includegraphics[width=58mm]{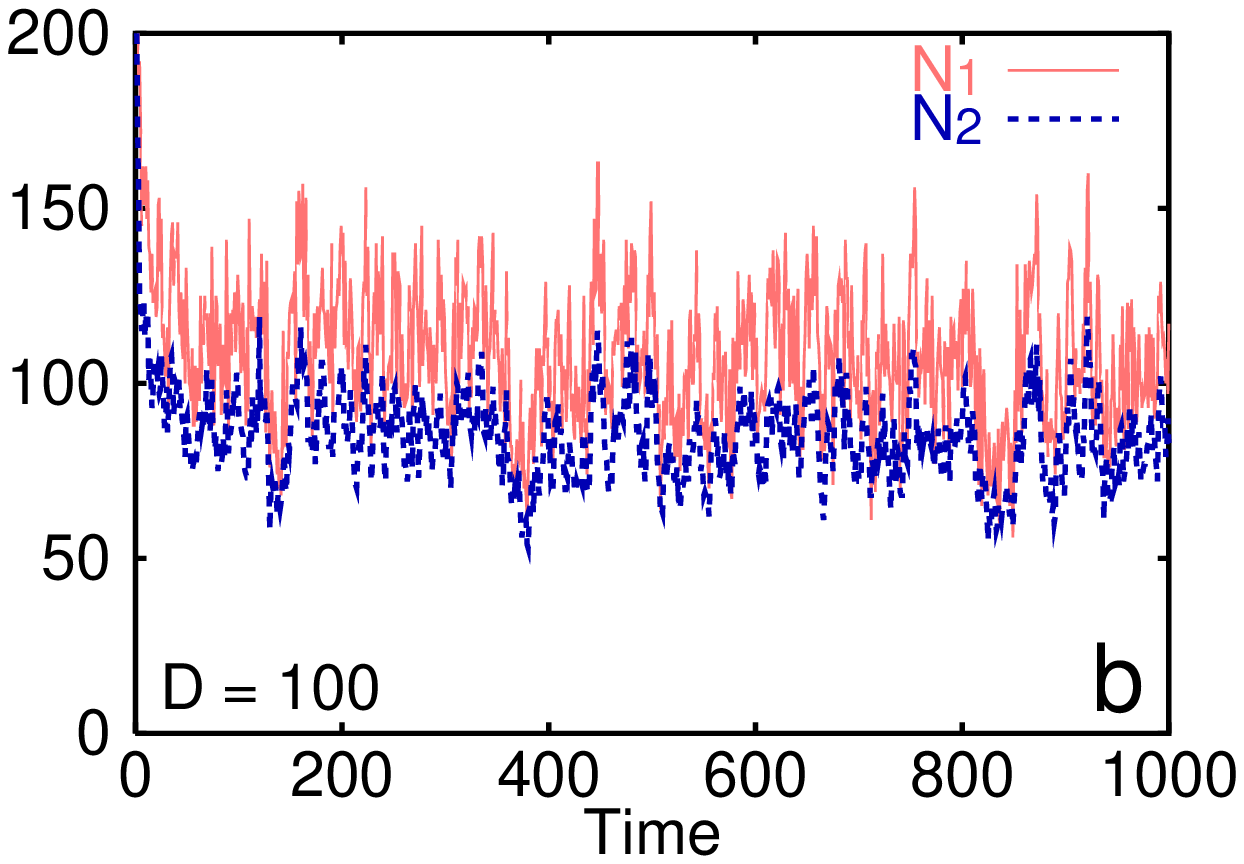}
\includegraphics[width=58mm]{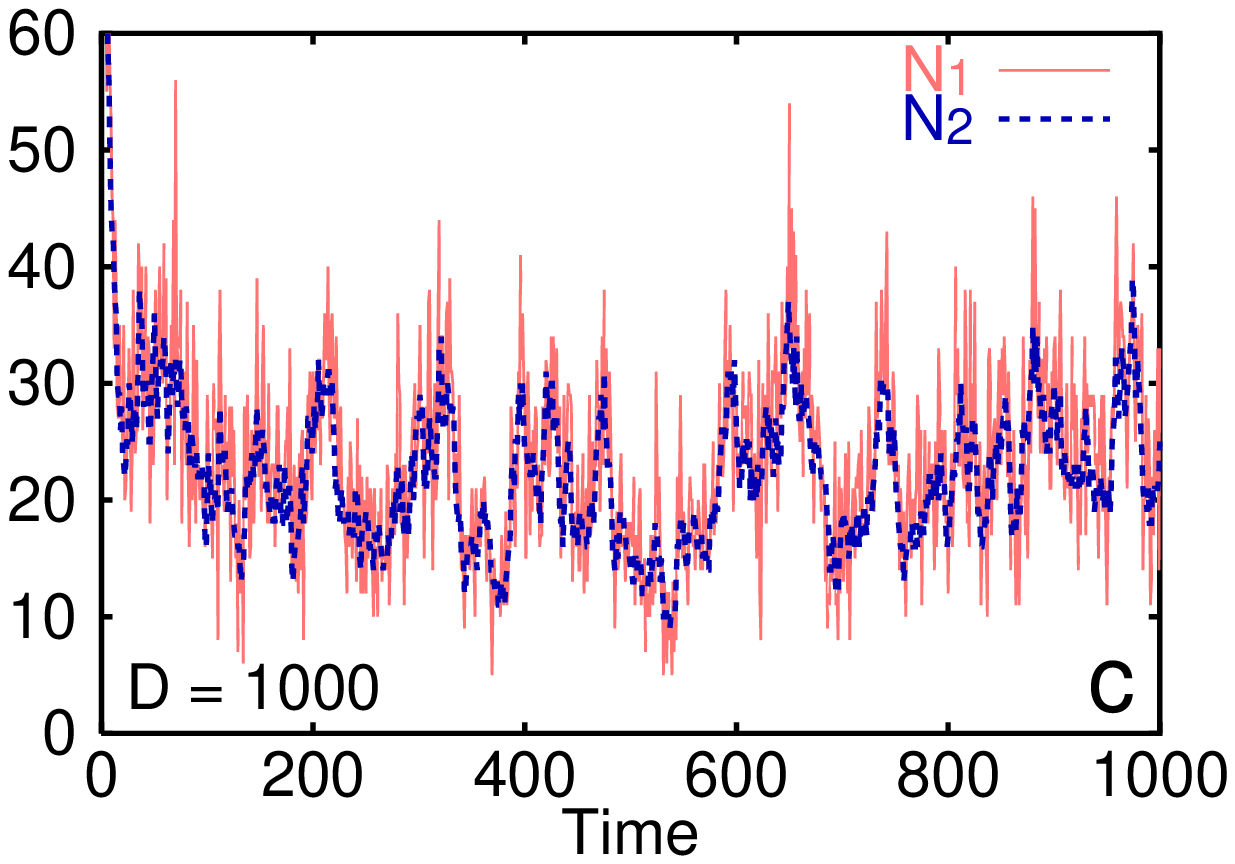}
\end{center}
\caption{(Color online)
Time series of $N_{1}$ and $N_{2}$.
$r=1$, $a = 4$, $N=1000$, $L_{x}=1000$.
a) $D=10$, b) $D=100$, c) $D=1000$.
Initially, $(N_{1}, N_{2}, N_{3}) = (250, 250, 500)$.
For $D=10$, $X_{3}$ reaches $0$, which corresponds to the unstable fixed
point $(2c/3, c/3, 0)$.}
\label{fig-sp1-1-ts-d}
\end{figure*}

\begin{figure}
\begin{center}
\includegraphics[width=80mm]{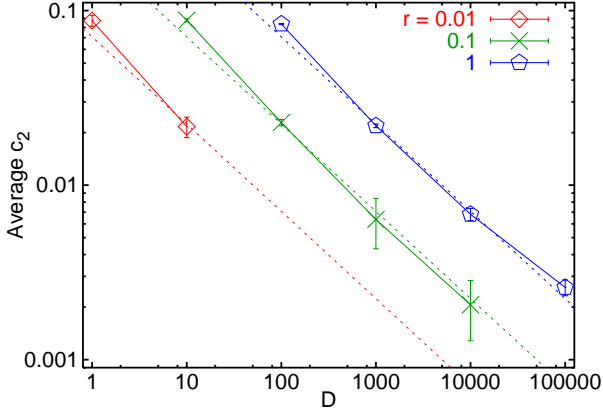}
\end{center}
\caption{(Color online)
Average concentration of $X_{2}$,
for different $r$ and $D$ ($a = 4$, $N=1000$, $L_{x}=1000$,
sampled over $5000 < t <10000$, and 10 trials.
The error bars show the standard deviation between the trials).
The dotted lines correspond to $0.1$ molecule per the Kuramoto length
$l_{1} = \sqrt{D / 50r}$ for each $r$.}
\label{fig-sp1-1-nbylk-d}
\end{figure}

\begin{figure}
\begin{center}
\includegraphics[width=80mm]{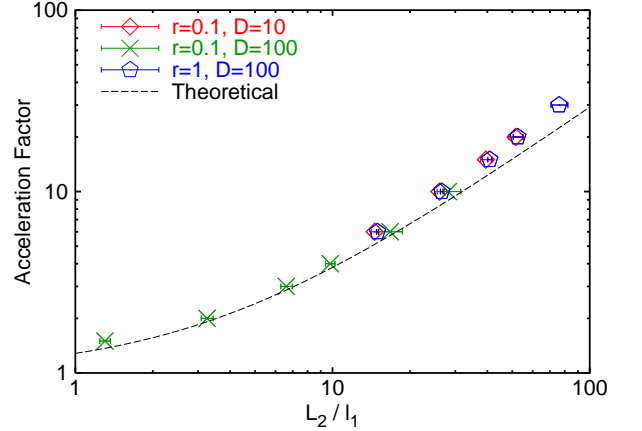}
\end{center}
\caption{(Color online)
The acceleration factor $\alpha$, plotted against $\lambda_{2} / l_{1}$.
We measure the relation from simulations
with different $r$, $D$, and $a$ ($N=1000$, $L_{x}=1000$, sampled over
$5000 < t <10000$, and 10 trials.
The error bars show the standard deviation of $c_{2}$ between the trials).
This is very close to the theoretical estimation
$\alpha = 1 + \frac{1}{2\sqrt{\pi}} \cdot \frac{\lambda_{2}}{l_{1}}$.
}
\label{fig-sp1-1-accel}
\end{figure}

Many systems in nature that involve chemical reactions can be studied
with the help of reaction-diffusion equations.
%
% Just an example but perhaps you can think of a better on.
%
For certain processes, a relatively small number of suitably chosen
continuous macroscopic variables yields excellent descriptive results.
In biological systems, however, not only is the variety of chemicals enormous,
the number of molecules of each of the chemical species can range from the
relatively very large to the relatively very small.
Now, if the species with small numbers of molecules were irrelevant,
obviously, their existence could be ignored and one could focus on the species
with large numbers of molecules that can effectively be described by a
continuous variable.
However, it should not really come as a surprise that it was found that,
in general, species with small numbers of molecules cannot be neglected and
that certain functions in cells can critically depend on very small
fluctuations \cite{Mikhailov,Blumenfeld}.
Indeed, in prior studies on reaction-diffusion systems some effects of
fluctuations on pattern formation were found
(see e.g., \cite{fluctuation1,fluctuation2}).
Stochastic differential equations are often used to study effects of fluctuations.

Of course, on a microscopic level chemicals are composed of molecules,
and the actual reactions occur between these molecules.
Therefore, in principle, reaction events must be integer and
change only discretely.
In analysis with stochastic differential equations, though,
the fluctuations are regarded as continuous changes.
Clearly, this approximation can only be valid if applied to fluctuations
that involve sufficiently large numbers of molecules and should not be applied
when relevant chemical species are very rare.

In order to address this issue, we previously studied the effects of discreteness
in simple autocatalytic reaction network systems and reported discreteness-induced
transitions as well as drastic effects on concentrations \cite{YTKK2001,YTKK2003}.
A key feature of these systems was, however,
that the medium was assumed to be well-stirred.

In contrast, in a system with diffusion in space, the total number of
molecules may vary from point to point.
By assuming that the reaction is fast and the diffusion is slow,
locally, the discreteness of the molecules can become important.
In fact, this can even be the case if the total number of molecules is
large but spread out over a large area as well.

Therefore, a length scale should be considered such that it can serve as
a benchmark for judging whether or not a continuum approximation is applicable.
To consider this problem, the ratio between the reaction and diffusion rates
is important and a candidate for the length scale is the
typical distance over which a molecule diffuses during its lifetime,
i.e., before it undergoes reaction
as defined by Kuramoto \cite{Kuramoto1,Kuramoto2}.
For reference, let us briefly review the work.

Consider the reaction
\footnote{In \cite{Kuramoto1,Kuramoto2}, this is expressed as
$A+M \rightarrow X+M$, $2X \rightarrow E+D$,
where, the concentrations of $A$, $M$, $E$ and $D$ are taken to be
constant in order to make an analysis of the equations possible.}
\[
A \stackrel{k}{\longrightarrow} X,\quad 2X \stackrel{k'}{\longrightarrow} B.
\]
If the concentration of $A$ is set to be constant,
 $X$ is produced at a constant rate $k$
while decaying by the reaction $2X \rightarrow B$ at a rate $k'$.
The average concentration of $X$ at the steady state is
$\langle X \rangle = \sqrt{kA/2k'}$,
where, for simplicity, $A$ is the concentration of the chemical $A$.
%
% the above is a bit silly "A is the concentration of A"
%
Thus the average lifetime of $X$ at the steady state is estimated to be
$\tau = 1 / (2k'\langle X \rangle) = 1 / \sqrt{2kk'A}$.
Suppose that $X$ molecules diffuse with the diffusion constant $D$.
The typical length over which an $X$ molecule diffuses in its lifetime
is then estimated to be
\begin{equation}
l = \sqrt{2D\tau},
\label{eqn:kuramoto}
\end{equation}
which is called the Kuramoto length \cite{Kampen}.

The Kuramoto length $l$ represents the relation
between the reaction rate and the diffusion rate.
When the system size is smaller than $l$,
its behavior is dominated by diffusion and
local fluctuations rapidly spread throughout the system.
Contrastingly, if the system size is much larger than $l$,
 fluctuations are localized only in a small part of the system,
and distant regions fluctuate independently.

In this reasoning, it is assumed that the average distance between
molecules is much smaller than $l$.  Thus the actual discreteness of
the molecules can be ignored, and the concentration of the chemical
$X$ can be
regarded as a continuous variable.
However, if the average distance
between molecules is comparable to or larger than $l$, local
discreteness of molecules may not be negligible.  Suppose a chemical
$A$, with very low concentration, produces another chemical $B$.  The
average lifetime of $B$ is short, such that the Kuramoto length of $B$
is shorter than the average distance between adjacent $A$ molecules.
With this setting, chemical $B$ molecules may be considered as localized
around $A$ molecules.
This is especially so if the  reactions involve 2nd or higher orders of $B$.
Then the localization
of chemical $B$ may drastically alter the total rate of the reactions,
and the effect of the local discreteness of the molecules
may thus be rather significant.

In order to systematically investigate the effects of the local discreteness
of the molecules, we consider a simple one-dimensional
reaction-diffusion system
with 3 chemicals ($X_{1}$, $X_{2}$, and $X_{3}$) and
the following 4 reactions
\begin{eqnarray*}
X_{2} + X_{3} \stackrel{k_{1}}{\longrightarrow} X_{2} + X_{1};\quad X_{3} +
X_{1} \stackrel{k_{2}}{\longrightarrow} 2 X_{3}\\
2 X_{2} \stackrel{k_{3}}{\longrightarrow} X_{2} + X_{1};\quad 2 X_{1}
\stackrel{k_{4}}{\longrightarrow} X_{1} + X_{2}.
\end{eqnarray*}
Here, we assume that the first two reactions are much faster
than the others, i.e., the reaction constants satisfy
$k_{1}, k_{2} \gg k_{3} > k_{4}$.
To be specific, we take
$k_{1}=k_{2}=100r$, $k_{3}=a r$, and $k_{4}=r$ ($r > 0$, $1 < a \ll 100$).

In the continuum limit, $c_{i}(t,x)$,
the concentration of chemical $X_{i}$ at time $t$ and position $x$,
is governed by the reaction-diffusion equation for the system
given by
\begin{eqnarray}
\frac{\partial c_{1}}{\partial t} & = & -100r(c_{1}-c_{2})c_{3} -
r(c_{1}^{2} - a c_{2}^{2}) + D_{1} \frac{\partial^{2} c_{1}}{\partial
x^{2}}\ \label{eqn:rd1}\\
\frac{\partial c_{2}}{\partial t} & = & r(c_{1}^{2} - a c_{2}^{2}) + D_{2}
\frac{\partial^{2} c_{2}}{\partial x^{2}} \label{eqn:rd2}\\
\frac{\partial c_{3}}{\partial t} & = & 100r(c_{1}-c_{2})c_{3} + D_{3}
\frac{\partial^{2} c_{3}}{\partial x^{2}} \label{eqn:rd3}
\end{eqnarray}
where $D_{i}$ is the diffusion constant of $X_{i}$.
The system is closed and thus the total concentration $c$ is conserved.
For simplicity, we assume $D_{i} = D$ for all $i$.

The reaction-diffusion equation has fixed points
at $(c_{1},c_{2},c_{3}) = (0, 0, c), (\sqrt{a}c/(\sqrt{a} + 1), c/(\sqrt{a}
+ 1), 0)$ for all $x$.
By performing a straightforward linear stability analysis, it is shown that
only the former is stable. Indeed,
by starting from an initial condition with $c_{i} > 0$,
this reaction-diffusion equation always converges to the fixed point $(0, 0,
c)$.

Next, in order to obtain insights into the case when the continuum limit
cannot be taken we carry out direct particle simulations.
Each molecule diffuses randomly (showing Brownian motion)
in a one-dimensional space
with periodic boundary conditions (length $L_{x}$).
When two molecules are within a distance $d_{r}$ they react
with a certain probability and
the total number of molecules ($N$) is conserved.

First, we investigate the case with $a = 4$ and show
time series of the number of molecules $N_{i}$ of chemical species $X_{i}$
in Fig. \ref{fig-sp1-1-ts-d}. As can be seen,
$N_{1}$ and $N_{2}$ do not converge to $0$ but to relatively large numbers.
As can be expected
the final concentrations depend on $r$ and $D$ and for $X_{2}$ it is depicted in
 Fig. \ref{fig-sp1-1-nbylk-d}.
Approximately, the concentration turns out to be proportional
to $\sqrt{r/D}$ when $N_{1}, N_{2} \ll N$.

To elucidate the origin of this proportionality,
we take a closer look at the Kuramoto length,
which, of course, depends on the molecule species.
In the case of the $X_{1}$ molecules it is
given by $l_{1} = \sqrt{D/50rc_{3}}$,
as the average lifetime of $X_{1}$ is $1/100rc_{3}$.
Here we consider the situation $N_{1}, N_{2} \ll N$,
so that $c_{3} \approx c$. In the discussion below, we assume that
$l_{1} = \sqrt{D/50rc} = \sqrt{DL_{x}/50rN}$.

Using this length $l_{1}$, the density of the remaining $X_{2}$ molecules
is found to be about $0.1$ molecule per $l_{1}$,
independent of the parameters,
as shown in Fig. \ref{fig-sp1-1-nbylk-d}.
After relaxation, this density does not depend on the initial conditions,
as long as $N_{i} \gg 1$ is satisfied initially.
Furthermore, the density is independent of the system size
$L_{x}$, if $L_{x} \gg l_{1}$,
so that the number of remaining molecules $N_{2}$ is simply proportional to
$L_{x}$.
Consequently, in this analysis one obtains a finite $c_2$ regardless of the
system size or initial conditions which is
clearly different from the continuum limit where $c_2$ goes to
$0$.

In this system, $X_{1}$ molecules are produced by $X_{2}$ molecules.
If $\lambda_{2}$, the average distance between $X_{2}$ molecules,
is smaller than $l_{1}$, the distributions of $X_{1}$ around neighboring
$X_{2}$ molecules overlap each other significantly and one can regard
 $X_{1}$ to be uniformly distributed.
In contrast, if $\lambda_{2}$ is
much larger than $l_{1}$, molecules $X_{1}$ will localize around the
$X_{2}$ molecules (The size $L_{x} \gg \lambda_{2}$).
Then, the reaction $2 X_{1} \rightarrow X_{1} + X_{2}$
is accelerated when compared to the case that the same total number of $X_{1}$
molecules is uniformly distributed.

We define the acceleration factor $\alpha(\lambda_{2},l_{1})$
as the ratio between the reaction rate with localized $X_{1}$
and the reaction rate with uniformly distributed $X_{1}$.
If $\lambda_{2} \gg l_{1}$, it is expected that $\alpha \gg 1$.
Assuming that the distribution of $X_{1}$ is continuous and
represented by the concentration $c_{1}(x)$
\footnote{Here, only the $X_{1}$ species is relevant to this reaction, so
that
it is not necessary to consider detailed structures smaller than the
typical distance between $X_{1}$ molecules and
 the total rate of the reaction can therefore be described by a
smoothened distribution.},
the acceleration factor can be expressed as
\begin{equation}
\alpha = \frac{\langle c_{1}^{2} \rangle}{\langle c_{1} \rangle^{2}} =
\frac{L_{x}^{-1}\int c_{1}^{2} dx}{\left(L_{x}^{-1} \int c_{1}
dx\right)^{2}}.
\label{eqn:alpha0}
\end{equation}

For simplicity, we assume that the distribution of the
localized $X_{1}$ molecules is
Gaussian with a standard deviation $l_{1}$ centered around the $X_{2}$ molecules
(which may overlap each other).
Suppose that the $X_{2}$ molecules are randomly distributed over the system
with an average distance $\lambda_{2}$,
we then obtain \footnote{The acceleration factor $\alpha$ is estimated as
follows.
We assume that the distribution of localized $X_{1}$ molecules is
Gaussian with a standard deviation $l_{1}$ around the $X_{2}$ molecules. I.e.\
 $\rho_{i}(x) = (\sqrt{2\pi}l_{1})^{-1} \exp(-(x-x_{i})^{2} /
2l_{1}^{2})$,
where $x_{i}$ is the position of each $X_{2}$ molecule.
The total distribution (concentration) of $X_{1}$ is $c_{1}(x) = \sum_{i}
\rho_{i}(x)$,
and $\langle c_{1} \rangle = \int \rho_{i}(x) dx / \lambda_{2} = 1 /
\lambda_{2}$.
Since the molecules $X_{2}$ are randomly distributed,
$\langle c_{1}^{2} \rangle = \left\langle (\sum \rho_{i})^{2} \right\rangle\\
= \left( \sum \langle \rho_{i} \rangle \right)^{2} + \sum \langle \rho_{i}^{2} \rangle
= \langle c_{1} \rangle^{2} + (2\sqrt{\pi}l_{1})^{-1} \langle c_{1} \rangle$\\
($L_{x} \gg l_{1}, \lambda_{2}$).
Thus,
$\alpha = \langle c_{1}^{2} \rangle / \langle c_{1} \rangle^{2}\\
= 1 + (2\sqrt{\pi}l_{1})^{-1} \langle c_{1}\rangle^{-1}
= 1 + \lambda_{2} / (2\sqrt{\pi}l_{1})$.\\
Consequently, we obtain $\displaystyle \alpha = 1 + \frac{1}{2 \sqrt{\pi}}
\cdot \frac{\lambda_{2}}{l_{1}}$.
}
\begin{equation}
\alpha = 1 + \frac{1}{2 \sqrt{\pi}} \cdot \frac{\lambda_{2}}{l_{1}}
= 1 + \frac{1}{2 \sqrt{\pi} \cdot l_{1} c_{2}}.
\label{eqn:alpha}
\end{equation}

On the other hand, the average lifetime of $X_{2}$ molecules is much longer,
so that the Kuramoto length for $X_{2}$ molecules is longer than $\lambda_{2}$.
%
% Above you just said that X_2 is randomly distributed so I cut out the sentence
% but now probably part of the argument is missing.
%
Consequently, the reaction $2 X_{2} \rightarrow X_{2} + X_{1}$ is not accelerated
by localization.

Provided that $N_{1}, N_{2} \ll N_{3}$,
$N_{1} \approx N_{2}$ due to the fast reactions
$X_{2} + X_{3} \rightarrow X_{2} + X_{1}$ and
$X_{3} + X_{1} \rightarrow 2 X_{3}$.
As a result, the ratio between  the two reaction rates is given by
\begin{equation}
\frac{\textrm{The rate of } (X_{1} \rightarrow X_{2})}{\textrm{The rate of }
(X_{2} \rightarrow X_{1})}
\approx \frac{\alpha k_{4} N_{1}^{2}}{k_{3} N_{2}^2}
\approx \frac{\alpha}{a}.
\label{eqn:rratio}
\end{equation}

Following eq. (\ref{eqn:rratio}), the two reaction rates are balanced
if $N_{2}$ takes a value such that $\alpha = a$ is satisfied.
Corresponding to $\alpha = a$,
a novel fixed point appears at
\begin{equation}
c_{1} = c_{2} = \left(2(a - 1)\sqrt{\pi}l_{1}\right)^{-1} (= c_{s}),
\label{eqn:fixedpoint}
\end{equation}
provided $c_{1}, c_{2} \ll c_{3}$ and $c_{3} = c$.
The stability of this fixed point is analyzed, by linearizing
eqs. (\ref{eqn:alpha}) and (\ref{eqn:fixedpoint}) around the fixed point.
Noting that
\begin{equation}
\alpha = 1 + \frac{(a - 1) c_{s}}{c_{2}} = a - \frac{a - 1}{c_{s}}\delta
c_{2} + o(\delta c_{2}),
\label{eqn:alphafp}
\end{equation}
with $c_{1} = c_{s} + \delta c_{1}$ and $c_{2} = c_{s} + \delta c_{2}$, and
rewriting eqs. (\ref{eqn:rd1}) and (\ref{eqn:rd2}) with $\alpha$ in eq.
(\ref{eqn:alphafp}),
we obtain
\begin{eqnarray}
\left(
\begin{array}{c}
\dot{c_{1}}\\
\dot{c_{2}}
\end{array}
\right)
\!\!&=&\!
r \left(
\begin{array}{cc}
- 2a c_{s} - 100c & (3a - 1) c_{s} + 100c\\
2a c_{s} & - (3a - 1) c_{s}
\end{array}
\right)\!
\left(
\begin{array}{c}
\delta c_{1}\\
\delta c_{2}
\end{array}
\right) \nonumber
\\
 & & \ + o(\delta c_{1}, \delta c_{2}).
\label{eqn:fplinear}
\end{eqnarray}
The Jacobi matrix has two negative eigenvalues,
and the fixed point is stable (This is natural, since if $\alpha <a$,
$N_2$ decreases, leading to
the increase of $\alpha$, and vice versa).
This fixed point (steady state) is distinct from that of
the original reaction-diffusion equation, $(0,0,c)$.

From eq. (\ref{eqn:alpha}),
$\alpha$ becomes $4$ when $\lambda_{2} / l_{1} = 6\sqrt{\pi} \approx 10.6$.
In our simulation with $a = 4$,
about $0.1$ $X_{2}$ molecule per $l_{1}$ remains,
as shown in Fig. \ref{fig-sp1-1-nbylk-d}.
In other words, $\lambda_{2} / l_{1} \approx 10$,
in good agreement with the estimate.

By changing $a$, we numerically obtain
the relation between the $\lambda_{2} / l_{1}$ and
the actual acceleration factor $\alpha$,
again agreeing well with the above theoretical estimate
$\alpha = 1 + \frac{1}{2 \sqrt{\pi}} \cdot \frac{\lambda_{2}}{l_{1}}$,
as shown in Fig. \ref{fig-sp1-1-accel}.

In the estimate above, we consider the case that $N_{1}, N_{2} \ll N$.
On the other hand, if $N$ is set to be
smaller than the estimated value of $N_{2}$ at the steady state,
$N_{2}$ increases to satisfy the balance, and finally reaches the state
$N_{1}+N_{2}=N$, $N_{3}=0$, which corresponds to the unstable fixed point
of the reaction-diffusion equation,
$(\sqrt{a}c/(\sqrt{a} + 1), c/(\sqrt{a} + 1), 0)$,
as shown in Fig. \ref{fig-sp1-1-ts-d} (a).

The localization of $X_{1}$ cannot be maintained without the spatial
discreteness of $X_{2}$ molecules.
In reaction-diffusion equations,
any pattern will disappear eventually given a sufficiently long evolution
time unless it is somehow sustained.
This is even the case when the initial distribution of $X_{2}$ is discrete.
But again, it is essential to recall that reaction-diffusion equations are
an approximation and in that sense an idealization.
In reality, a single molecule itself can of course not be broadened by diffusion
and the spatial discreteness of $X_{2}$ molecules is always maintained.
By itself, a molecule is a diffusion-resistant pattern.

The alteration of the steady state due to localization is not limited to
the present type of reaction network. Provided that the conditions
\begin{enumerate}
\item[(i)] Chemical $A$ generates another chemical species $B$.
\item[(ii)] The lifetime of $B$ is short or the diffusion of $B$ is slow
so that the Kuramoto length of $B$ is much smaller
than the average distance between $A$ molecules.
\item[(iii)] The localization of molecule $B$ accelerates some
reactions.
\end{enumerate}
are satisfied, discreteness may alter the dynamics.
The last condition is
easily satisfied if species $B$ is involved in second or higher order reactions.
Finally, if
\begin{enumerate}
\item[(iv)] The acceleration alters the density of $A$ molecules,
\end{enumerate}
the above acceleration mechanism may control the density of $A$ to produce a
novel steady state.

As for the localization effect by the discreteness of catalytic molecules,
Shnerb \textit{et al.} recently showed that it can amplify autocatalytic
reaction-diffusion processes \cite{Solomon2000,Louzoun}.
In their model, however, the density of the catalyst is fixed as an
externally given value, and the concentration of the product,
localized around the catalyst, diverges in time.
In our mechanism, the density of the catalyst ($A$, or $X_{2}$)
changes autonomously and reaches a suitable value to produce the discreteness
effect.
Hence the effect of discreteness is controlled by the discreteness itself,
leading to a novel steady state.
Indeed, theoretical estimates for the novel concentrations
based on the self-consistent fixed point of acceleration due to the
localization agree well with numerical results.

In so far as the conditions (i)--(iv) are met,
our result does not depend on the details of the reactions,
and should generally be valid for reaction-diffusion systems.
% For purposes of verification of the presented results,
We have carried out simulations of similar reaction-diffusion systems,
and again the discreteness effect led to novel pattern formation
that cannot be accounted for by Turing type mechanisms (with or without noise).

Experimental verification of our
results should be possible by
suitably designing a reaction system, with the use of, say,
microreactors or vesicles.
Also, in biological cells, many chemicals work at low concentrations
on the order of 1 nM or less. Furthermore,
diffusion is sometimes restricted, e.g.\ due to surrounding macro-molecules,
and may be slow.
In such an environment, it is probable that the average distance
between the molecules of a given chemical species is much larger
than the Kuramoto lengths of some of the other chemical species.
Indeed, biochemical systems contain various higher order reactions
%(for example, catalyzed by enzyme complexes)
and positive feedback mechanisms
that might naturally support the conditions (iii)--(iv) above.

% Acknowledgements
\begin{acknowledgments}
This research is supported by grants-in-aid for scientific research from
the Ministry of Education, Culture, Sports, Science and Technology of Japan
(11CE2006, 15-11161).
One of the authors (Y.T.) is supported by a research fellowship from
Japan Society for the Promotion of Science.
\end{acknowledgments}

%--------


\begin{thebibliography}{99}
\bibitem{Kampen}
N. G. van Kampen, \textit{Stochastic Processes in Physics and Chemistry}
(North-Holland, rev. ed., 1992),
where, to be precise,
the length is defined as the half of $l$, but the difference
does not affect the discussion here.
\end{thebibliography}
\end{document}